\documentclass[12pt]{article}
\usepackage{times}
\usepackage{graphicx}
\usepackage{cite}
\usepackage{hyperref}
\usepackage{amsmath}
\usepackage{amssymb}
\usepackage[font=small,labelfont=bf]{caption}
\usepackage{xcolor}
\usepackage{ulem}

\title{Thermal properties of the superconductor-quantum Hall interfaces}

\author
{Lingfei Zhao$^{1\dagger\ast}$, Trevyn F.Q. Larson$^{1\dagger}$, Zubair Iftikhar$^{1\dagger}$, John Chiles$^{1}$,
\\ Kenji Watanabe$^3$, Takashi Taniguchi$^3$, Fran\c cois Amet$^2$ and
Gleb Finkelstein$^{1\ast}$
\\
\normalsize{${}^{1}$Department of Physics, Duke University, Durham, NC 27708, USA}\\
\normalsize{${}^{2}$Department of Physics and Astronomy, Appalachian State University, Boone, NC 28607, USA}\\
\normalsize{${}^{3}$National Institute for Materials Science, 1-1 Namiki, Tsukuba 305-0044, Japan}\\\normalsize{$\dagger$These authors contributed equally.}\\
\normalsize{$\ast$Corresponding author. E-mail: lz117@duke.edu (L.Z.); gleb@duke.edu (G.F.)}
}

\date{}

\begin{document} 

\baselineskip 15pt

\maketitle 

\begin{quote} 
{\bf 
An important route of engineering topological states and excitations is to combine superconductors (SC) with the quantum Hall (QH) effect, and over the past decade, significant progress has been made in this direction. 
While typical measurements of these states focus on electronic properties, little attention has been paid to the accompanying thermal responses. Here, we examine the thermal properties of the interface between a type-II superconducting electrodes and graphene in the QH regime. We use the thermal noise measurement to probe the local electron temperature of the biased interface. Surprisingly, the measured temperature raise indicates that the superconductor provides a significant thermal conductivity, which is linear in temperature. This suggests electronic heat transport and may be unexpected, because the number of the quasiparticles in the superconductor should be exponentially suppressed. Instead, we attribute the measured electronic heat conductivity to the overlap of the normal states in the vortex cores.
}
\end{quote}

Over the past decade, significant progress has been made in combining quantum Hall states with superconductors~\cite{Amet2016,lee_inducing_2017,Seredinski2019,Zhao2020,Gl2022,Vignaud2023}.
These and other experiments clearly demonstrate the coherent aspects of the transport along the proximitized channels.
However, little experimental evidence exists for the thermal properties of the QH-SC interfaces. Due to the poor heat conductivity of the semiconductors and superconductors, 
even a seemingly small heat dissipation at their interface could result in a significant rise of the local temperature~\cite{Denisov2022}, destroying the delicate effects induced by superconducting proximity. In this paper, we explore the heat balance in the hybrid superconductor-quantum Hall structure. 

Thermal properties of the nanoscale quantum materials emerged as an interesting novel area of research~\cite{Pekola_RevModPhys}. While such quantities as Nernst and Seebeck coefficients became relatively well known, thermal conductivity measurements remain less explored~\cite{Waissman2021}. Recent studies demonstrate that measurements of mK temperatures and fW powers are feasible in appropriately designed QH samples via Johnson-Nyquist noise~\cite{Jezouin2013,Melcer2023}.

We use a similar noise setup to examine the thermal response of a hybrid device made of graphene contacted by type-II thin film superconducting electrodes. In the QH regime, we bias the device forming hot spots -- regions where the Joule heating is deposited -- at the interface with the superconductor, from where the heat can escape via several mechanisms. We then measure the noise carried by the QH edge states downstream of the superconductor to probe its local temperature. 

We find that applying current on the scale of tens of nA can heat the interface to $\sim 1$ K. As the magnetic field increases, the electron temperature gradually decreases until becoming comparable to the temperature of a similarly heated normal contact. We argue that this increase of the cooling efficiency is explained by the electronic heat conductivity of the superconductor, mediated by the normal states in the vortex cores~\cite{Golubov}. This mechanism becomes more efficient as the distance between the vortices decreases with magnetic field. Further examination of the temperature response across a few $\mu$m-wide superconducting strip reveals that indeed the heat spreads rather efficiently the across superconductor. Eventually, at high magnetic field, the whole width of the strip becomes uniformly heated. The estimated thermal conductivity varies by almost two orders of magnitude over the studied range of magnetic field.

\section*{Experimental setup}

\begin{figure}
\centering
\includegraphics[width=1\textwidth]{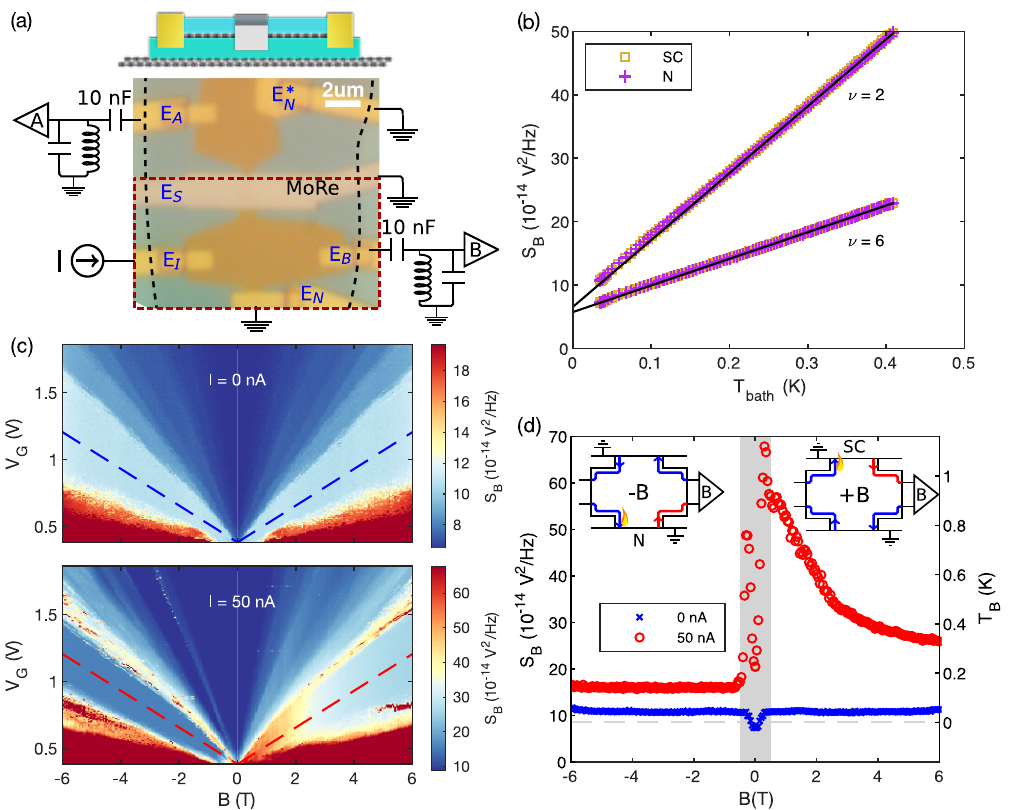}
\caption{\textbf{Excess noise of superconducting and normal contacts.}
(a) Schematics of the device and measurement configuration. Normal contacts (Cr/Au) are yellow and the superconducting contact (MoRe) is gray. The cross section (top) shows the hBN/graphene/hBN stack sitting on a graphite gate (outlined by black dashed lines in the optical image). The two-stage amplifier chains (A and B) are attached to the normal electrodes through LC resonators and 10 nF blocking capacitors. The LC resonator is composed of a stray capacitance $\sim$ 170 pF and an inductor $\sim$ 66 $\mu$H, giving a center frequency of 1.5 MHz. The superconductor, the bottom and top-right normal contacts are cold grounded. Current bias $I$ is injected at the bottom left normal contact. In the remainder of the figure, only the bottom region of graphene is measured.
(b) Zero-bias $S_\mathrm{B}$ of the normal metal (purple crosses, $B=$ --3 T) and superconductor (yellow boxes, $B=$ 3 T) plotted vs. the bath temperature $T_\mathrm{bath}$. The top and bottom curves are obtained on the $\nu=2$ and 6 plateaus. 
(c) Noise power $S_\mathrm{B}$ measured by amplifier B (after the amplification chain) plotted vs. $B$ and $V_G$ at zero bias (top) and 50 nA (bottom). 
(d) Cuts of the maps in (c) measured along the dashed lines (middle of $\nu=2$ plateau). Insets: schematics of the edge state direction, and locations of the relevant hot spots and hot edges for negative and positive fields. 
}
\label{fig1}
\end{figure}

The schematic and the image of our sample are shown in Figure~\ref{fig1}a. The device is made of graphene encapsulated in hBN (the thickness of the top/bottom layers: $\sim$ 30 and 60 nm) and placed on top of a graphite gate. The graphene crystal is cut into two separate regions by a superconducting contact, made of MoRe alloy (50-50 in weight, 80 nm thick). The lengths of the top and bottom interfaces are respectively 0.5 and 1 $\mu$m. 
Several additional normal contacts to graphene are made of a normal metal (90 nm of Au on 1 nm Cr).

Throughout the measurement, we apply a field of several Tesla to induce the QH effect in graphene. Both MoRe and Cr/Au form better contacts to the n-doped graphene, and we focus exclusively on the electron doping (positive gate voltages). In this case the chiral direction of the QH edges is clockwise (counterclockwise) for positive (negative) $B$. The device is held at a base temperature of 35 mK. 

The MoRe has a critical temperature $T_c\sim\ $10 K and stays superconducting at least up to 12 T, the highest field in this measurement (Figure~\ref{MoRe}).
The width of the superconducting strip (light gray in Figure~\ref{fig1}a) is $\sim$ 2 $\mu$m, much longer than the MoRe coherence length ($\xi<$ 10 nm). Therefore, in the first part of this paper the two graphene-superconductor interfaces will be considered separately. 

The noise at the two interfaces is measured by two homemade cryogenic amplifiers ($\sim$ 10 dB gain), A and B, which are attached to the normal electrodes, E$_\mathrm{A}$ and E$_\mathrm{B}$, at the top and bottom regions respectively (Figure~\ref{fig1}a). The amplifiers are AC coupled to the sample electrodes via 10 nF blocking capacitors and LC resonators centered around 1.5 MHz. The outputs of the cryogenic amplifiers are then fed to room temperature amplifiers (46 dB gain) and digitized. 

We start by exploring the thermalization of the edge state in contact with the superconductor. The QH-superconductor interface is expected to have a limited cooling power, and first we have to verify that it is not overheated by spurious currents. We work with the bottom region of the sample (Figure~\ref{fig1}a), and measure the 
spectral density $S_\mathrm{B}$ of amplifier B. Depending on the field direction, the QH edge channels will arrive at the amplifier contacts either from the superconducting strip (positive field) or from the bottom normal contact (negative field). To enable a direct comparison, the length of the bottom normal metal and the bottom superconducting interfaces are designed to be the same (1 $\mu$m), and both contacts are cold grounded.

Figure~\ref{fig1}b plots the zero-bias $S_\mathrm{B}$ measured vs. bath temperature $T_\mathrm{bath}$ for both $\nu=2$ and 6 at $|B|=$ 3 T. The result for the superconductor (squares) and the normal contact (crosses) are practically indistinguishable. The noise is linear vs. $T_\mathrm{bath}$, allowing us to calibrate the gain of the amplification chain and convert $S_\mathrm{B}$ to temperature. 
Next, we measure $S_\mathrm{B}$ at the base temperature $T_\mathrm{bath}=T_0$ as a function of $V_G$ and $B$ (top panel of Figure~\ref{fig1}c). The noise map is nearly symmetric with respect to the field direction, and the noise stays constant on the plateaus, indicating that both the superconducting and the reference normal contact stay thermalized at the base temperature of the sample, $T_0=35$ mK, throughout the full range of magnetic field. In the following, we disregard a narrow range of fields $|B|<1$~T, where the conductance of the sample is no longer quantized (see Figure ~\ref{DC}).

\section*{Excess noise}
We next apply a current bias of $I=50$ nA to the bottom-left contact and plot the resulting $S_B(V_G,B)$ in the bottom panel of Figure~\ref{fig1}c.
At negative $B$, when we locally heat the bottom normal contact, the noise on the $\nu=2$ plateau increases compared to zero bias. This indicates that the normal electrode is locally heated by the applied power, and its thermal noise is then detected by the amplifier B located downstream. Note that in this case, the excess noise  stays constant in $B$.
At positive $B$, when the superconductor is locally heated, the excess noise on the $\nu=2$ plateau stays constant in $V_G$, but strongly dependent on magnetic field. 

In Figure~\ref{fig1}d we plot the noise measured along the dashed lines corresponding to the $\nu=2$ plateaus in Figure~\ref{fig1}c. This noise is converted to the temperature via the calibration developed in Figure~\ref{fig1}b, and both the noise and the temperature scales are shown. At $I=0$, the noise on the $\nu=2$ plateau stays constant in 
$B$, except for the small vicinity of zero field. At finite bias, the elevated noise stays constant at negative fields, indicating that the normal contact is heated to about 100 mK independent of the field. At positive field corresponding to biasing the superconductor, we observe a strong enhancement of the noise followed by a gradual decrease. 

\begin{figure}
\centering
\includegraphics[width=0.9\textwidth]{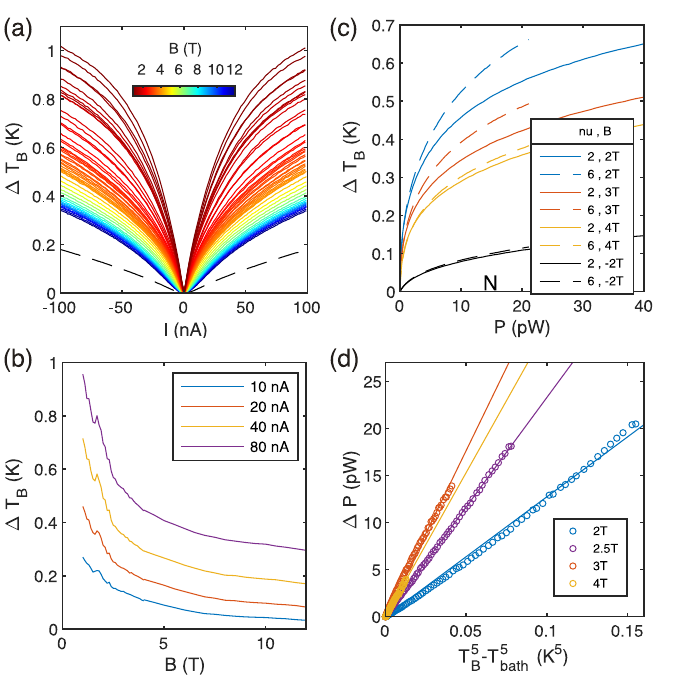}
\caption{\textbf{Heating of the bottom interface.} 
(a) Temperature increase $\Delta T_\mathrm{B}$ plotted vs. bias current $I$ for $\nu=2$. The different curves corresponds to magnetic fields $B$ is stepped by 0.1 T (1--4 T range) and 0.5 T (4 -- 12 T range). 
(b) $\Delta T_\mathrm{B}$ vs. $B$ for $\nu=2$ is plotted at constant bias currents of 10, 20, 40 and 80 nA, showing the gradual decay of temperature with field. Variations of temperature observed at small fields for all curves are likely caused by changes of the cooling pathways due to vortex rearrangements, see Figure~\ref{fluctuations}.   
(c) $\Delta T_\mathrm{B}$ plotted vs. the Joule power $P$ at $T_\mathrm{bath}=$ 35 mK. Different colors correspond to magnetic fields of 2, 3 \& 4 T, while $\nu=2$ and 6 are represented by the solid and dashed lines. The corresponding $\Delta T_\mathrm{B} (P)$ curves of the normal interface are nearly independent on $B$ and $\nu$ (black lines). 
(d) The difference in power between the $\nu=2$ and 6 curves of panel (c) plotted against $T_\mathrm{B}^5-T_\mathrm{bath}^5$. The linear dependence indicates that the difference can be attributed to the contribution of phonons.
}
\label{fig2}
\end{figure}

In principle, the increased noise at the biased superconductor-QH interface could have originated from the shot noise~\cite{Sahu2021}. Namely, in the case of an ideal superconductor, an incoming electron would have to undergo either a normal or Andreev reflection. Strong current fluctuations could then be expected, corresponding to the electrons or holes being emitted downstream (toward the amplifier). However, the probabilities of the normal and Andreev processes strongly depend on the gate voltage~\cite{Zhao2020}, while experimentally, the noise at a fixed field stays flat across the plateaus (Figure~\ref{fig1}c). (Very small variations of the noise, on the scale of 1\%, have been observed as a function of gate voltage.) Furthermore, our previous studies of the superconductor-QH interfaces, including this very sample~\cite{Zhao2023}, indicate that most of the electrons arriving at the superconducting contact are absorbed, likely by the Caroli-de Gennes-Matricon (CdGM) states of the superconducting vortices~\cite{Caroli1964}. We argue that as a result, the interfacial region of the superconductor is locally heated, and its thermal noise is carried downstream to the amplifier, similar to the case of the normal contact. We will provide further evidence of the thermal nature of the noise in Figures~\ref{fig2} and \ref{fig3}. Finally, we independently verified the temperature rise at the superconductor-QH interface by measuring the suppression of the non-local resistance pattern, qualitatively confirming the results of the noise thermometry (Figure~\ref{T_CAES}).

Noise $S$ measured by an amplifier is proportional to the average temperature of the incoming and outgoing QH channels at the amplifier contact. When a hot spot with a temperature $T$ is created at the upstream contact, the channels originating from the normal contact are expected to stay close to the bath temperature $T_0$. We then expect $S \propto (T_0+T)/2$, as compared to $S \propto T$ obtained in the case when the whole sample is uniformly heated. In the following, we convert the excess noise to the local electron temperature using the calibration of Figure~\ref{fig1}b. 

\section*{Cooling of the biased interface}

In Figure~\ref{fig2}a, we plot the temperature of the bottom superconducting interface $T(I)$, as the field is stepped from $B=1$ to 12 T. The gate voltage is adjusted to stay in the middle of the $\nu=2$ plateau. At $B=1$ T, the $1\mu$m-long superconductor-graphene interface reaches $\sim 1$ K at 100 nA. Even at a 10 nA bias (corresponding to a voltage drop of $\sim 130 \mu$V), the electron temperature increases to about 300 mK. As $B$ increases, $T$ decreases precipitously, as show by the $T(B)$ graphs measured at fixed $I$ in Figure~\ref{fig2}b. At the highest field of 12 T, the temperature of the superconducting interface becomes comparable to that of the reference normal contact, whose $T(I)$ is plotted in Figure~\ref{fig2}a as a dashed line. 
Interestingly, at that point MoRe is still superconducting (Figure~\ref{MoRe}).

The $T(I)$ curves in Figure~\ref{fig2}a have a V-shape typical for hot electrons in a metal that are cooled via diffusion (Wiedemann–Franz mechanism) and the emission of phonons~\cite{Jezouin2013}.  Namely, at low temperatures, the phonon emission is negligible, and the applied power $P\propto I^2$ is balanced by electronic cooling $\propto (T^2-T_0^2)$, resulting in the roughly linear slope of the $T(I)$ curves visible in Figure~\ref{fig2}a. At higher temperatures, the emission of phonons by hot electrons results in the sublinear $T(I)$~\cite{Huard2007}.

\section*{The role of phonons}

Next, we plot $T(P)$ -- the temperature of the interface as a function of applied power for three magnetic field values (2, 3 \& 4 T) in Figure~\ref{fig2}c. Here $P=I^2R/2$, where  the factor of $1/2$ appears because the Joule heating is evenly distributed between two hot spots. For comparison, we also plot the $T(P)$ of the normal metal at 2 T, which are nearly independent of $B$. 

The focus of this figure is to compare the behavior for the filling factors $\nu=2$ (solid lines) and $\nu=6$ (dashed lines). At a given field, the pairs of the $T(P)$ curves for $\nu=2$ and 6 overlap at low power, but diverge at high power. Surprisingly, the temperature at $\nu=6$ is hotter, ruling out the more efficient cooling via QH edges. Moreover, the difference is noticeable at higher power, where the phonon emission is the dominant cooling mechanism. To prove this point, we plot the difference in power, $\Delta P$, needed to reach the same $T$ vs. $T^\delta-T_0^\delta$ and find a linear relation for $\delta=5$ (Figure~\ref{fig2}d). $\Delta P \propto (T^5-T_0^5)$ is the typical cooling power of the hot electrons via phonon emission in diffusive metals~\cite{Huard2007}.  

While we cannot pinpoint the exact origin of the difference between the two filling factors, we attribute it to the difference in the size of the QH hot spot, which is determined by the length over which the QH edge channel is equilibrated with the contact. Notice that a similar difference is also observed for the normal contact (black lines) -- $\nu=2$ is slightly cooler than $\nu=6$ -- so this effect is not specific to the superconductor.

Finally, we use the data in Figure~\ref{fig2} to bring forward two arguments further supporting the attribution of the excess noise to the electron temperature. First, $T(P)$ curves for $\nu=2$ and 6 nearly coincide at low power. However, a significantly enhanced shot noise could have been expected for $\nu=6$ vs. $\nu=2$, because a given power $P$ would correspond to a $\sqrt 3$ times higher bias current. Second, in principle we could still attempt to fit the $S(I)$ curves with the shot noise expression, $S \propto I \times \textrm{coth} (eV/2k_B T_e)$, with $V=hI/\nu e^2$. The curvature at the minimum of the $S(I)$ curves would then correspond to the electrons' base temperature $T_e$ treated as a fitting parameter. $T_e$ would then depend on both $\nu$ and $B$, increasing with magnetic field to $\sim 200$ mK at 12 T. This result would both be physically unreasonable and contradict Figure~\ref{fig1}d. We therefore rule out the shot noise as an explanation of the measured $S(I)$ curves.

\section*{The role of electrons}

We can expect that phonon emission should be negligible for electron temperatures below about 100 mK. Namely, we estimate the corresponding cooling power as $P_{ph}=\Sigma V (T^5-T_0^5)$, where $\Sigma \sim 1$nW $\mu$m$^{-3}$K$^{-5}$ is the typical coupling constant ~\cite{Huard2007}, and $V$ is the volume of the metal, which can be at most $0.1 - 1\mu m^3$ in our case. In fact, the effective volume should be even smaller, as normal electron behavior is only expected for the CdGM states which occupy a small part of the superconducting film. The resulting cooling power should be at most $\sim 1-10$ fW at 100 mK, which is negligible compared to the hot electron diffusion, as we show next. 

\begin{figure}
\centering
\includegraphics[width=1\textwidth]{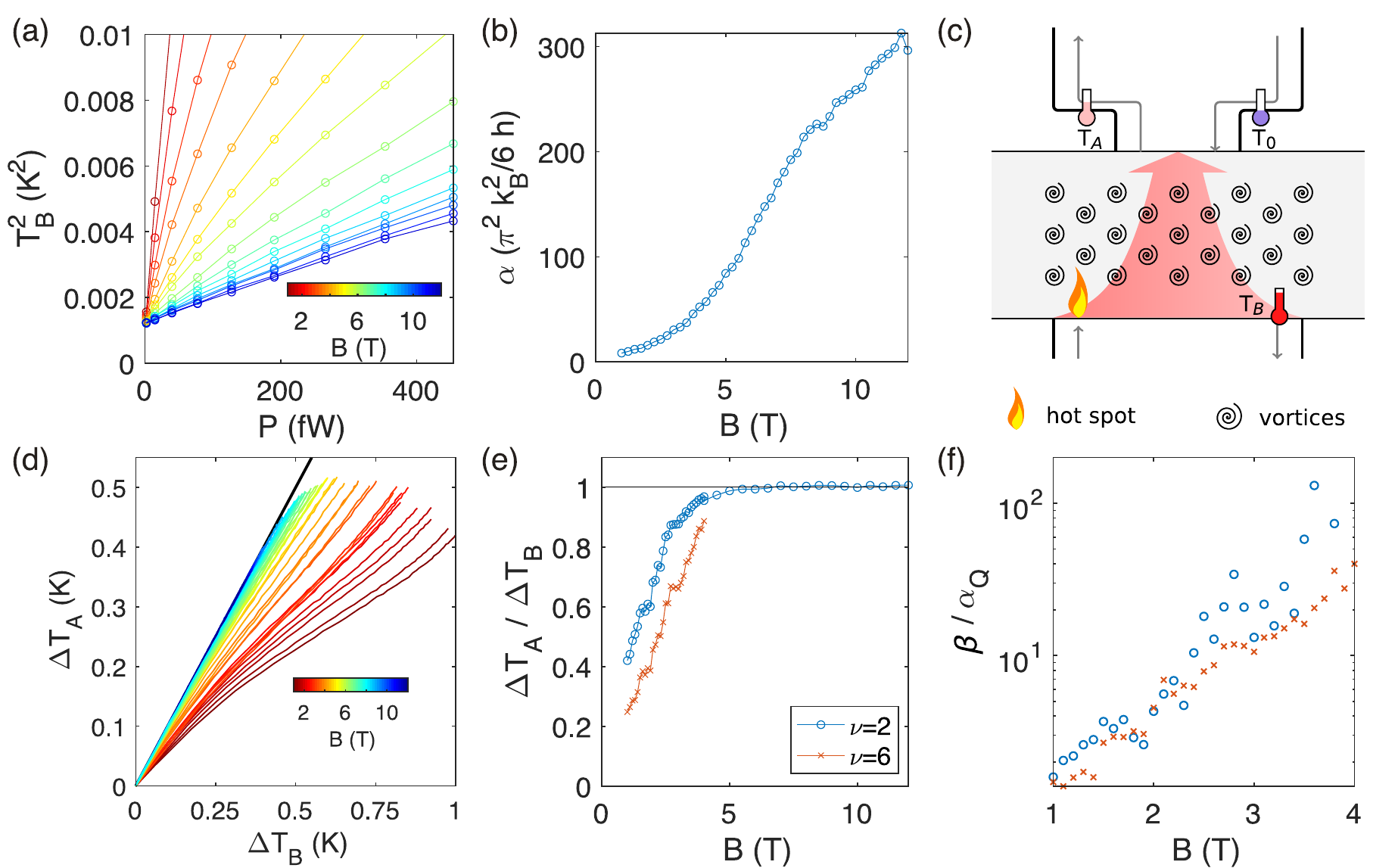}
\caption{\textbf{Thermal conductivity of the superconducting film.}
(a) $T_\mathrm{B}^2$ plotted vs. Joule heating power $P$ from $B=$1 to 11.5 T (shown in steps of 0.75 T). Linear dependence is clearly visible indicating electronic thermal conductivity.
(b) The slope $\alpha$ extracted from the $T_\mathrm{B}^2(P)$ curves in panel (a) as a function of $B$. $\alpha$ is represented in units of $\alpha_Q = \pi^2 k_B/ 6h$, corresponding to a single QH channel.
(c) Schematics of the heat flow across the superconductor showing the heat balance at the top interface. 
(d) $\Delta T_\mathrm{A}$ plotted vs. $\Delta T_\mathrm{B}$ on the $\nu=2$ plateau from $B=$ 1 to 12 T. The $T_\mathrm{B}$ data is the same as in Figure~2a. The black line is $\Delta T_\mathrm{A}=\Delta T_\mathrm{B}$.
(e) $\Delta T_\mathrm{A}/\Delta T_\mathrm{B}$ ratio extracted from the lower range of temperatures in panel (d), plotted vs. $B$ for $\nu=2$ (blue circles) and 6 (red crosses). The quantity converge to unity at high magnetic field. 
(f) $\beta$, plotted in units of $\alpha_{Q}$, represents the electronic thermal conductivity across the top interface (please see main text). $\beta(B)$ dependencies nearly overlap for $\nu=2$ (blue circles) and $\nu=6$ (red crosses). 
}
\label{fig3}
\end{figure}

To further analyze the purely electronic cooling, in Figure~\ref{fig3}a we plot $T^2$ vs. applied power in the range of $P< 500$ pW and $T<100$ mK . The data points show a clear linear dependence, indicating that the superconducting interface is indeed cooled via the diffusion of hot electrons and the emission of phonons can be neglected, with a possible exception of the lowest fields ($B \lesssim 1.5$ T). We further extract the slope of the $P = \alpha (T_\mathrm{B}^2-T_0^2)$ dependence (Figure~\ref{fig3}b). The coefficient $\alpha$ is presented in the quantized units of $\alpha_Q = \pi^2 k_B/ 6h$, corresponding to the heat flow in a single QH channel~\cite{Jezouin2013}. Even at the lowest field, the measured $\alpha$ greatly exceeds that of the quantum Hall. We therefore attribute it to the electronic heat conductivity of the superconductor, which appears to be linear in temperature, like in a normal metal.

At zero field, the electronic heat conductivity of a superconductor is strongly suppressed due to the gap in the single-particle spectrum. (Though violation of this dependence has been reported see e.g.~\cite{Feshchenko2017}.) However, at finite field, the CdGM states should behave like a normal metal, and the tunneling of the quasipartciels between the vortex cores can efficiently conduct heat. The resulting thermal conductivity is linear in temperature and rapidly grows with magnetic field due to the increasing overlap between the vortex cores~\cite{Golubov}. This regime was tentatively reported in the seventies~\cite{Vinen1971}, and more recently explored in Ref.~\cite{Boaknin2003}. Closer to $H_{c2}$, the order parameter becomes nearly uniformly suppressed, which increases the density of quasiparticles~\cite{Caroli1965}. In this regime, $\alpha$ is linearly approaching the value corresponding to the normal state of the metal~\cite{Willis_perpfield}. Both regimes --  the initial rapid rise followed by flattening close to $H_{c2}$ -- are clearly visible in Figure~\ref{fig3}b, which may be the first such demonstration in thin film samples.

\section*{Temperature spread across the superconductor}
We are now interested to find out what happens once the heat spreads from the QH hot spot through the superconductor. To that end, we use amplifier A (Figure~\ref{fig1}a) to study the temperature of the top superconducting interface, $T_\mathrm{A}$. (We also reinstate subscript B to indicate the temperature of the bottom interface, $T_\mathrm{B}$.) In Figure~\ref{T_A_vs_P}, we plot $T_\mathrm{A}$ vs. $P$ while the power is still applied by biasing the bottom left contact, the same way as in Figure~\ref{fig2}. Evidently, the heat spreads efficiently to the opposite side of the superconducting strip. 

In Figure~\ref{fig3}d we plot $\Delta T_\mathrm{A} = T_\mathrm{A}-T_0$ vs. $\Delta T_\mathrm{B}$. Experimentally, we find that $\Delta T_\mathrm{A}$ scales almost linearly with $\Delta T_\mathrm{B}$. The robust nature of this relation, which extends over the range of temperature changes much larger than the base temperature, is presently not clear. We extract the  $\Delta T_\mathrm{A}/\Delta T_\mathrm{B}$ slope at low power and plot it in Figure~\ref{fig3}e as a function of $B$. Similar data for several applied powers is presented in Figure~\ref{TATB3I}. Independent of $P$, the slope increases monotonically with $B$ and saturates at 1 for $B \gtrsim 4$ T, directly confirming that the whole width of the superconducting strip gets uniformly hot. 

To understand the behavior in the lower field range, $B \lesssim 4$T, we note the difference between the plots of $\Delta T_\mathrm{A} / \Delta T_\mathrm{B}$ measured at $\nu=2$ and $\nu=6$  (lower curve in Figure~\ref{fig3}e). We know that the superconductor is highly thermally conductive for the whole range of fields $B>1$ T (Figure~\ref{fig3}b), and the 2 vs. 6 additional quantum channels of graphene should not change the temperature of the metal itself. We therefore argue that $T_\mathrm{A}$ is the temperature of the QH edge running along the relatively short top interface, which has not fully equilibrated with the hot electrons in the superconductor. 

We analyze the heat flow at the top superconducting interface as sketched in Figure~\ref{fig3}c. Given the high heat conductivity of the superconductor (Figure~\ref{fig3}b), we approximate the electron temperature of the the superconducting metal close to the top interface as $T_\mathrm{B}$. The edge channels arrive at the interface with temperature $T_0$ and acquire heat from the superconductor, eventually reaching a temperature of $T_\mathrm{A}$. This temperature should satisfy the heat balance equation, $\beta(T_\mathrm{B}^2-T_\mathrm{A}^2)=\nu \alpha_{Q}\ (T_\mathrm{A}^2-T_0^2)$, where $\beta$ is proposed to represent the effective electronic heat conductivity across the top interface. Using this expression, we extract $\beta$ by fitting $T_\mathrm{B}^2-T_\mathrm{A}^2$ vs. $T_\mathrm{A}^2-T_0^2$ and find that the values nearly coincide for the two filling factors $\nu=2$ and 6 (Figure~\ref{fig3}f). We note that $\beta$ is different from $\alpha$ in Figure~\ref{fig3}b, which characterizes the electronic heat conductivity of the superconductor itself. In contrast, $\beta$ characterizes the heat conductivity between the metal and the edge state at the top interface.

\section*{Discussion}

Summarizing our results, we find that the dissipation at the hot spot results in a substantial overheating of the SC-QH interface which is particularly noticable at low magnetic fields. Since this is the regime most interesting for the design of topological states in hybrid superconducting structures, it's important to carefully engineer heat sinks so that fast gate operations for topological quantum computing can be achieved without overheating the electrons. 

Increasing the magnetic field gradually thermalizes the superconductor. Unintuitively, by $\sim H_{c2}/2$, the heat dissipation by the superconducting interface becomes comparable to that of the normal electrode. In this regime, the temperature of superconducting strip becomes uniform in the transverse direction. The change of the thermal conductivity by more than one order of magnitude (Figure~\ref{fig3}b) suggests that such films could potentially be used as a thermal switch when combined with ferromagnetic materials. 

Our original motivation for this study came from the expectation that probabilistic conversion of the chiral Andreev edge states to either electrons or holes could result in shot noise. However, only a small variation of the noise vs. $V_G$ on the scale of $\sim 0.01 eI $ was observed in one of the samples. Absorption of the particles by vortices enable multiple equilibration paths, which suppress the shot noise. Furthermore, our observations are  unfavorable for measuring the heat conductivity of the proximitized QH edge states along the SC interface -- unfortunately, they would be thermally shunted by the thermal conductivity of the superconductor. 

\section*{Acknowledgments}
We thank C. Beenakker for explaining the reason for the suppression of the shot noise in our measurement. G.F. appreciates the technical discussions of the amplification setup with M. Heiblum, F. Lafont, M. Reznikov, and Y. Ronen. The work at Duke University was supported by the Division of Materials Sciences and Engineering, Office of Basic Energy Sciences, U.S. Department of Energy, under Award No. DE-SC0002765. The deposition of MoRe was performed by F.A. at the Appalachian State University. K.W. and T.T. acknowledge support from the Elemental Strategy Initiative conducted by the MEXT, Japan, (grant no. JPMXP0112101001), JSPS KAKENHI (grant no. JP20H00354) and CREST (no. JPMJCR15F3, JST).
The sample fabrication was performed at the Duke University Shared Materials Instrumentation Facility (SMIF), a member of the North Carolina Research Triangle Nanotechnology Network.

\bibliography{HeatSC}
\bibliographystyle{naturemag}









\newpage
\section*{Supplementary Information}

\global\long\def\theequation{S\arabic{equation}}
\global\long\def\thefigure{S\arabic{figure}}
\setcounter{equation}{0}
\setcounter{figure}{0}

\subsection*{Superconducting transition of the film}
\begin{figure}[h]
\centering
\includegraphics[width=0.8\textwidth]{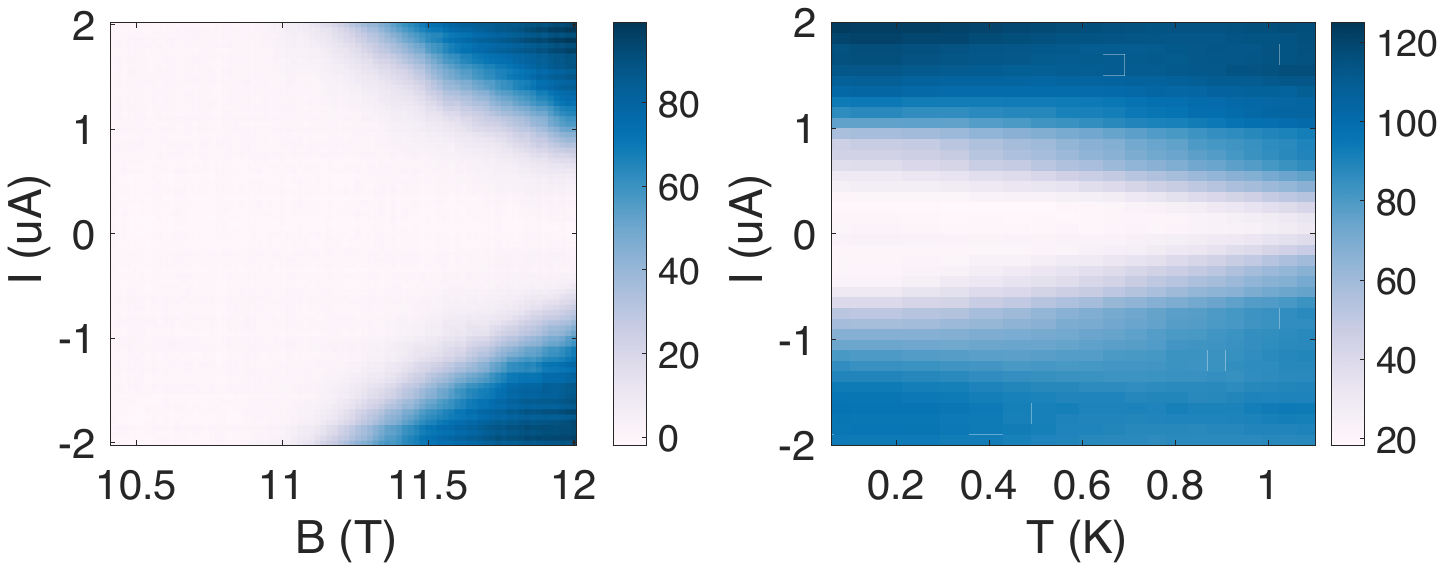}
\caption{Resistance (Ohms) of the superconductor strip plotted vs. $I$ and $B$ at the base temperature (left) and vs. $I$ and $T$ at 12 T (right).
} 
\label{MoRe}
\end{figure}

In Figure~\ref{MoRe}a, we present the  resistance of MoRe vs. current, magnetic field and temperature. Here, the current is applied along the strip forming the central contact in Figure~\ref{fig1}a. Evidently MoRe remains superconducting in the full range of the magnetic fields studied. Figure~\ref{MoRe}b shows that at 12 T, the superconductor strip has a critical current of $\sim 0.5 \mu$A and a critical temperature around 1 K. 

\subsection*{DC conductance}

\begin{figure}[h]
\centering
\includegraphics[width=0.8\textwidth]{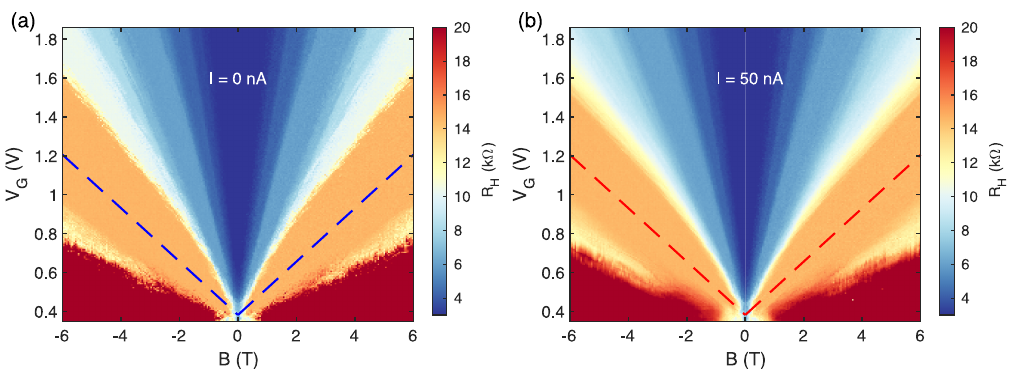}
\caption{DC conductance of the bottom region of the sample plotted vs. $V_G$ and $B$ at zero bias (a) and 50 nA bias (b).
} 
\label{DC}
\end{figure}

As a companion to the noise fan plotted in Fig.~\ref{fig1}c, we show the corresponding DC conductance of the bottom region in Fig.~\ref{DC}. The dashed line marks the center of $\nu=$ 2 plateau. 

\subsection*{Variations of cooling efficiency due to vortex rearrangement}
\begin{figure}[h]
\centering
\includegraphics[width=0.6\textwidth]{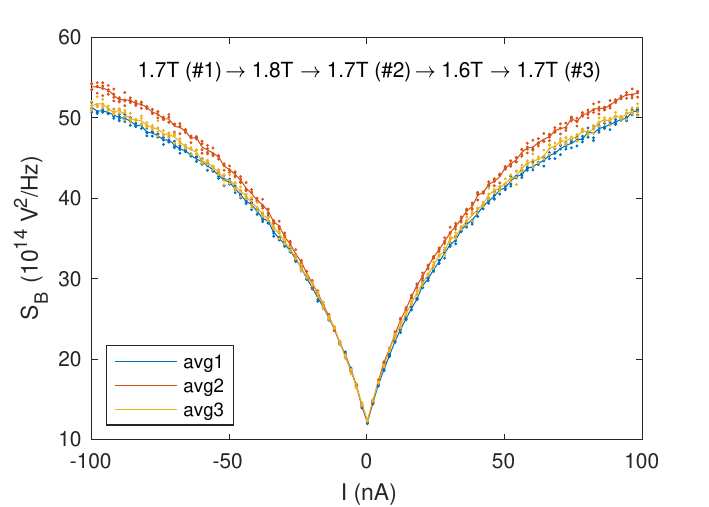}
\caption{Temperature increase in raw noise units $S_\mathrm{B}$ plotted versus bias current $I$ at $1.7$ T. Between each curve, the magnetic field is ramped by $+0.1$ or $-0.1$ T and then back to $1.7$ T. 
Each curve (colored solid line) is an average of 4 current bias sweeps, presented as solid scatter dots of the same color to show the measurement accuracy.
} 
\label{fluctuations}
\end{figure}

The temperature of the locally heated interface should depend on the positions of vortices that define the cooling paths for hot electrons. In turn, the vortex configuration depends not only on the value of the field but also on its history. In Figure~\ref{fluctuations}, we measure the $S_B(I)$ curve at 1.7 T three times, ramping the field up to 1.8 T between measurement \#1 and \#2, and ramping it down to 1.6 T between measurement \#2 and \#3. Our earlier study of the same sample shows that the change of field by 0.1 T is sufficient to significantly change the vortex configuration ~\cite{Zhao2023}. 

Each measurement in Figure~\ref{fluctuations} is composed of 4 sweeps of bias current (scattered dot symbols of the same color) that are then averaged (solid lines). One can see that at finite bias, the average temperature of measurement \#2 is well above the other two, beyond error bars. This indicates that the change of the vortex configuration  between the three sweeps results in differences between the cooling efficiency. As a result, the temperatures shown in the main text are affected by randomness induced by specific vortex configurations. 



The resulting fluctuations in the $T_\mathrm{A}(B)$ and $T_\mathrm{B}(B)$ data are prominently visible at low fields in Figures 2b and 3e and may appear as noise. However, based on Figure~\ref{fluctuations} we conclude that these fluctuations are due to non-deterministic vortex reconfigurations and not the measurement inaccuracy.


\newpage

\subsection*{Temperature of the top interface vs. power}

\begin{figure}[h]
\centering
\includegraphics[width=0.6\textwidth]{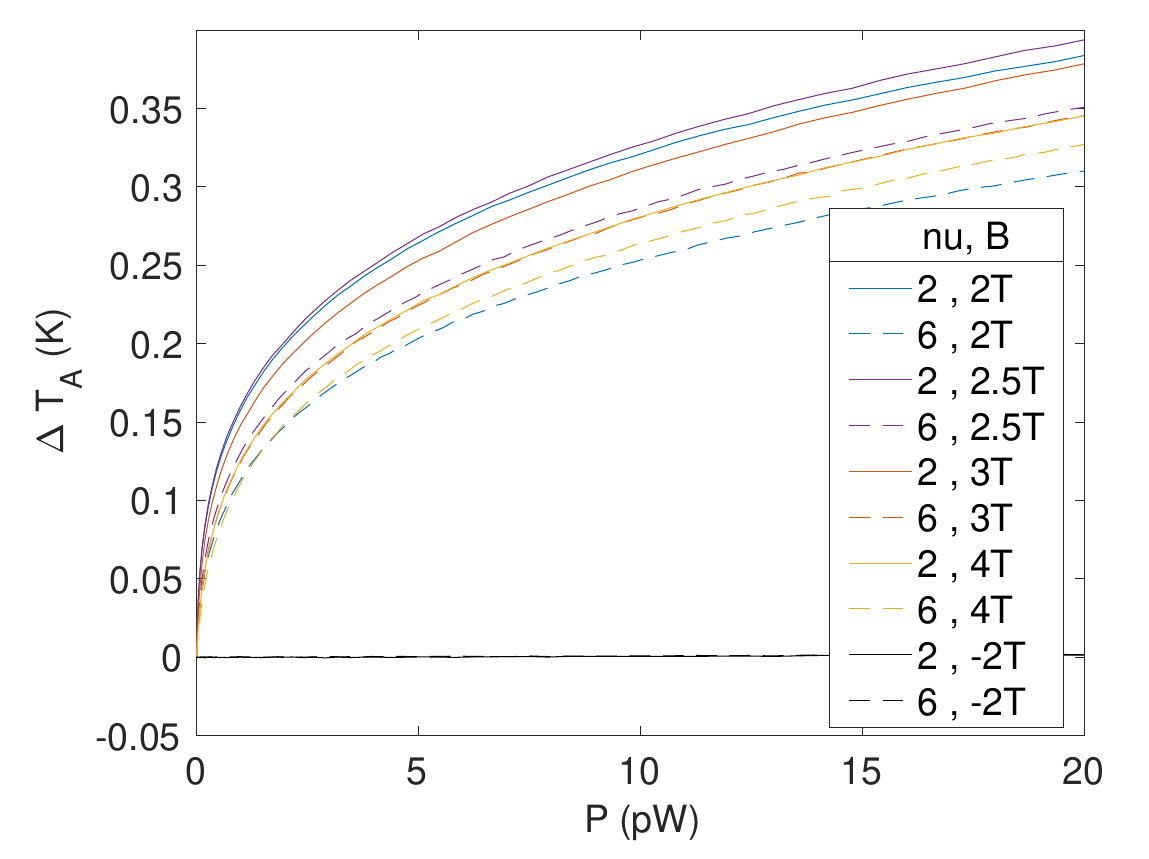}
\caption{Temperature of the top interface vs. power applied at the bottom part of the sample. The top curves correspond to positive field (heating the superconducting strip), while the nearly horizontal lines correspond to the negative field (heating only normal contacts).
} 
\label{T_A_vs_P}
\end{figure}

In Figure~\ref{T_A_vs_P}, we plot $\Delta T_\mathrm{A}$ vs. $P$ for $\nu=2$, 6 and several fields, similar to Figure~\ref{fig2}c. In positive fields, $\Delta T_\mathrm{A}$ grows dramatically with $P$, indicating that the temperature increase has spread from the bottom to the top superconducting interface. This observation serves as another confirmation that the excess noise is thermal -- no shot noise could be expected when biasing across a wide superconducting strip. 

As a check, we also plot $T_\mathrm{A}$ in negative field, corresponding to the opposite HQ chirality (black line close to the $x$ axis). In that case, the hot spots are formed at normal contacts away form the superconductor, and the rise of $T_\mathrm{A}$ is very small ($<$ 5 mK), which indicates that the heat transfer through the substrate is negligible.

\newpage
\subsection*{$\Delta T_\mathrm{A}/\Delta T_\mathrm{B}$ at different applied bias current}
\begin{figure}[h]
\centering
\includegraphics[width=0.8\textwidth]{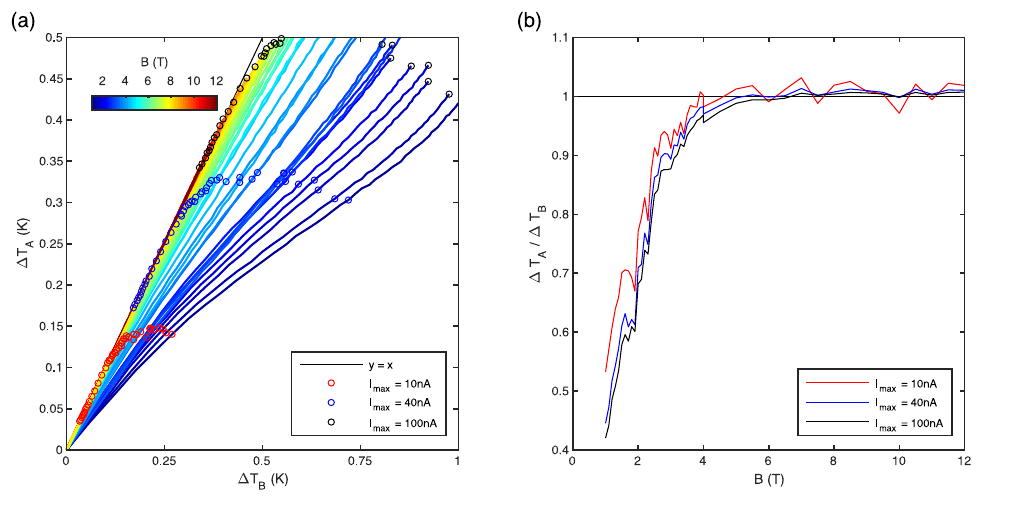}
\caption{
(a) A copy of Figure~\ref{fig3}d replotted with circles labeling the data points measured at an applied bias current of 10 nA (red), 40 nA (blue) and 100 nA (black). (b) The ratio of $\Delta T_\mathrm{A}/\Delta T_\mathrm{B}$ at these biases plotted vs. $B$.
} 
\label{TATB3I}
\end{figure}

In the main text, we noticed the nearly linear relation between $\Delta T_\mathrm{A}$ and $\Delta T_\mathrm{B}$ over a wide range of temperatures. Here, we replot Figure~\ref{fig3}d Figure~\ref{TATB3I}a. We extract the average slope for three different values of applied heating current and indicate the corresponding temperature range by a series of dots in Figure~\ref{TATB3I}a. The resulting $\Delta T_\mathrm{A}/\Delta T_\mathrm{B}$
is plotted in Figure~\ref{TATB3I}b, which shows that the curves are nearly independent of power. 

In particular, above $\sim 4$ T, $\Delta T_\mathrm{A}$ is very close to $\Delta T_\mathrm{B}$, independent of the applied power. This behavior is particularly surprising, because the applied power changes by a factor of a 100, from $0.5$ to $50$ pW. The corresponding $T_\mathrm{B}(P)$ dependence transitions from being determined by hot electron diffusion to phonon emission (Figure~\ref{fig2}c, Figure~\ref{fig3}a). In the latter regime, in order for $T_\mathrm{A}$ to reach $T_\mathrm{B}$, the heat should be carried across the superconducting strip by phonons and transferred to electrons, without leaking to the substrate. 


\newpage
\subsection*{CAES thermometry}

\begin{figure}[h]
\centering
\includegraphics[width=1\textwidth]{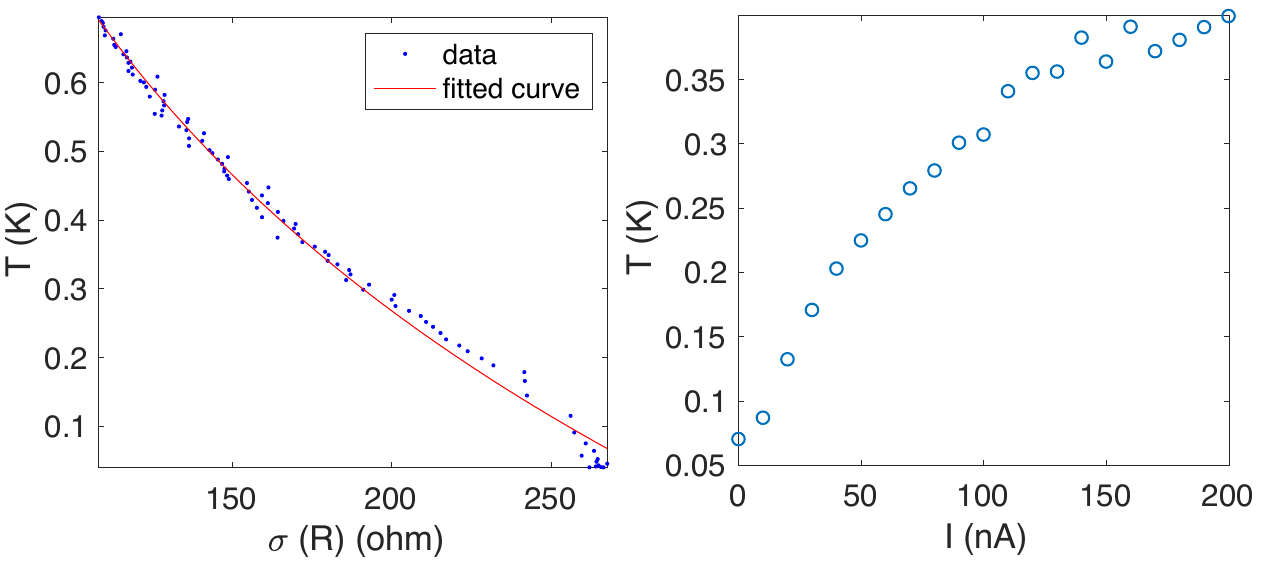}
\caption{CAES thermometry on the $\nu=2$ plateau at $B=$ 3 T. On the left, we plot the bath temperature $T$ vs. the corresponding zero-bias standard deviation of the CAES downstream resistances, $\sigma(R)$, of the top graphene-superconductor interface. The red curve is a linear fit of $T$ vs. $\log\left(\sigma(R)\right)$. On the right, we plot $T$ calibrated using the red curve as a function of the bias at the bottom interface. 
} 
\label{T_CAES}
\end{figure}

In our previous works, we have explored interference of chiral Andreev edge states (CAES) formed along the superconductor interface~\cite{Zhao2020}. We have shown that the resulting non-local ``downstream'' resistance is exponentially sensitive to the temperature~\cite{Zhao2023}. In this section, we use the downstream resistance of the top interface as a thermometer to conduct an independence check of the results presented in the main paper. This measurement is performed in a separate cooldown where we only connect DC lines to the sample. In the left panel of Figure~\ref{T_CAES}, we present the temperature ($y$-axis) vs. standard deviation of the downstream resistance of the top interface ($x$-axis). The latter quantity is measured at $B=3$ T across the $\nu=2$ plateau. A linear fit of $T$ -- $\log\left(\sigma(R)\right)$ is used as the calibration (red curve) of the resulting CAES thermometer. Then we measure $\sigma(R)$ at the base temperature as a function of bias current $I$ applied to the bottom interface. $\sigma(R)$ is then converted to the temperature of the top interface and plotted in the right panel. We indeed find a familiar sub-linear temperature rise on the same scale as the results obtained by the noise thermometry.

\end{document}